\documentclass[twocolumn,showpacs,pra,aps]{revtex4}
\usepackage{amssymb}
\usepackage{amsmath}
\usepackage{amsfonts}
\usepackage{graphics}
\usepackage{epic}
\usepackage{eepic}
\usepackage{color}
\usepackage{epsfig}
\usepackage{bbm}
\usepackage{graphicx}

\begin{document}

\def\ra{\rangle}
\def\la{\langle}
\def\bege{\begin{equation}
  }

\def\ende{\end{equation}}

\def\begarr{\begin{eqnarray}
  }

\def\endarr{\end{eqnarray}}
\def\ha{{\hat a}}
\def\hb{{\hat b}}
\def\hu{{\hat u}}
\def\hv{{\hat v}}
\def\hc{{\hat c}}
\def\hd{{\hat d}}
\def\no{\noindent}\def\non{\nonumber}
\def\hi{\hangindent=45pt}
\def\v{\vskip 12pt}

\newcommand{\bra}[1]{\left\langle #1 \right\vert}
\newcommand{\ket}[1]{\left\vert #1 \right\rangle}

\title{A Random Matrix Model of Adiabatic Quantum Computing}

\author{David R.\ Mitchell$^{1,2}$}
\author{Christoph Adami$^{1,4}$}
\author{Waynn Lue$^3$}
\author{Colin P.\ Williams$^{1,3}$}

\affiliation{
$^1$Jet Propulsion Laboratory, \\
California Institute of Technology, MS 126-347,
 4800 Oak Grove Drive, CA~91109-8099 \\
$^2$Northwestern State University, Natchitoches, Louisiana 71497\\
$^3$Stanford University, Stanford, CA 94305\\
$^4$Keck Graduate Institute, Claremont, CA 91711.}

\date{\today}


\begin{abstract}
We present an analysis of the quantum adiabatic algorithm for
solving hard instances of 3-SAT (an NP-complete problem) in terms
of Random Matrix Theory (RMT). We determine the global regularity
of the spectral fluctuations of the instantaneous Hamiltonians
encountered during the interpolation between the starting
Hamiltonians and the ones whose ground states encode the solutions
to the computational problems of interest.  At each interpolation
point, we quantify the {\em degree of regularity} of the average
spectral distribution via its {\em Brody parameter}, a measure
that distinguishes regular (i.e., Poissonian) from chaotic (i.e.,
Wigner-type) distributions of normalized nearest-neighbor
spacings. We find that for hard problem instances, i.e., those
having a {\em critical} ratio of clauses to variables, the
spectral fluctuations typically become irregular across a
contiguous region of the interpolation parameter, while the
spectrum is regular for easy instances. Within the hard region,
RMT may be applied to obtain a mathematical model of the
probability of avoided level crossings and concomitant failure
rate of the adiabatic algorithm due to non-adiabatic Landau-Zener
type transitions. Our model predicts that if the interpolation is
performed at a uniform rate, the average failure rate of the
quantum adiabatic algorithm, when averaged over hard problem
instances, scales exponentially with increasing problem size.

\end{abstract}

\maketitle

\section{Introduction}
Can quantum computers solve NP-complete prob lems
using physical resources of time, space, and energy that are all
bounded by {\em polynomials} in the size of the problem? Most
computer scientists are skeptical of such a
possibility~\cite{Bennettetal1997}. However, recently a more
physics-inspired perspective has arisen that is causing some to
rethink this question. In 2001, Farhi et al. presented a quantum
adiabatic algorithm for solving an NP-complete problem, and showed
via numerical simulations on a sequence of progressively larger
problem instances that the running time of this algorithm appears
to grow only as a polynomial in problem size~\cite{Farhietal2001}.
By contrast, all known classical algorithms for solving
NP-complete problems require a running time that scales
exponentially with problem size in the worst
case~\cite{GareyJohnson1979}. If the polynomial scaling of the
adiabatic algorithm is correct, this would represent a monumental
result for the field of quantum computing, as it would bring a
host of useful but hard computations within the domain of
computational tasks that can be performed exponentially faster on
quantum computers than classical ones. With such extraordinary
promise, it behooves us to understand the adiabatic algorithm in
full detail. Unfortunately, it has proven to be exceedingly
difficult to obtain analytic results on the scaling behavior of
the quantum adiabatic algorithm. Instead, for the most part,
researchers have relied upon numerical simulations of small
problem instances, typically involving 23 variables or
less~\cite{Farhietal2001}. By comparison, a modern-day classical
algorithm for solving 3-SAT can routinely solve instances
containing several thousand
variables~\cite{Selmanetal1992,LynceMarquesSilva2002}. It is
questionable whether numerical results based on 23-variable
simulations can be extrapolated reliably to 2000-variable
instances. Thus a more analytic approach is needed.

In this paper we develop such an analysis based on Random Matrix
Theory (RMT)~\cite{Wigner1967,Brodyetal1981,Mehta1991}, which is a
statistical description of complex quantum systems in which detailed
knowledge about particle interactions is abandoned in favor of a
description in terms of random interactions. Such a description is
typically found to be applicable to complex Hamiltonians without
fundamental symmetries, and with very large phase space (high
dimension). In this case, the level density $\rho(E)$ and the
nearest-neighbor spacings (NNS) distribution assume universal laws,
and only the {\em statistical} properties of the levels are of
interest. Such Hamiltonian systems are usually highly irregular,
disordered, and chaotic.

RMT has proven to be a successful method for predicting properties
of complex quantum systems that look superficially very different in
terms of their energy eigenspectra. By working with the
nearest-neighbor level spacing fluctuations rather than the raw
eigenspectra, deep similarities between apparently different
physical systems have been revealed, and several hard to calculate
properties, such as transition rates, have been determined. In
essence, a tractable model Hamiltonian can be used to make
predictions about an intractable one provided their energy spectra
can be described by similar NNS distributions.

In standard RMT, Hamiltonians are drawn from an ensemble of
orthogonal or unitary Gaussian matrices (the Gaussian orthogonal or
unitary ensembles, GOE or GUE), which corresponds to Hamiltonians
with interactions of all possible particle ranks $d$ ($d$-body
interactions). In other words, every element of Hilbert space is
assumed to be connected to every other by an interaction strength
that is given by a random number. The level density corresponding to
such a model is the celebrated ``semi-circle law" of
Wigner~\cite{Wigner1967}, which predicts a power dependence between
the density of states and energy. Physical complex quantum systems
(such as large atoms or nuclei) on the other hand, are better
described by interactions of rank two, that is, two-body
interactions at the most. Such Hamiltonians can be obtained from the
so-called``two-body random ensemble"~\cite{MonFrench1975} (or {\it
embedded} GOE) that well describes complex nuclear and atomic level
spectra, but gives rise to a Gaussian, rather than polynomial,
density of states. For Hamiltonians drawn from either a GOE or an
embedded GOE, the distribution of nearest-neighbor level separations
is unimodal with a long tail, known as the {\it Wigner}
distribution~\cite{Brodyetal1981,Mehta1991}. Such distributions are
typical for complex systems without symmetries, which results in
highly irregular energy spectra and chaotic dynamics. Conversely,
Hamiltonians corresponding to physical systems subject to symmetries
and conservation laws typically display {\em regular} energy spectra
and {\em Poissonian} nearest-neighbor level separations. Such
distributions fall off much faster than the Wigner distribution, and
decay monotonically. The {\em Brody} distribution interpolates
between those two distributions with a single Brody parameter $q$,
where the limit $q=0$ corresponds to the Poissonian limit while
$q=1$ gives the Wigner distribution.

A priori, the Hamiltonians arising in computational problems such as
$k$-SAT appear to have such a special structure that they are
unlikely to be described by random interaction matrices. We can
assess this by characterizing the irregularity of the NNS
distribution of the instantaneous Hamiltonian of the adiabatic
algorithm. If we find a Brody parameter close to zero, RMT cannot be
used, while a Brody parameter closer to one indicates that RMT
theory can predict global properties of the Hamiltonian dynamics
reliably. Note that the scaling of the level density itself
(polynomial for GOE, exponential for embedded GOE) is irrelevant for
this determination.

We determine the NNS distribution of the instantaneous Hamiltonians
solving 3-SAT problems by generating random soluble problem
instances (with exactly one solution) with a fixed ratio of clauses
to variables, determining for each the eigenvalue distribution, from
which the fluctuations can be obtained. In short, the results reveal
a systematic change in the spectral regularity of the instantaneous
Hamiltonians during the course of the adiabatic algorithm. In the
initial phase of the interpolation for especially hard problem
instances, the statistical NNS fluctuations conform to a regular,
Poisson-type distribution. Later in the interpolation, the
fluctuations conform to an irregular Wigner-type distribution
instead. We also find that irregular spectra only occur for
computationally {\em hard} problem instances.

In this paper, we predict the scaling of the failure rate of the
adiabatic algorithm for a fixed ratio of clauses to variables at a
given point in the interpolation process, as larger and larger
problem instances are considered. The adiabatic algorithm fails when
the system spontaneously transitions from its ground state into any
excited state. If we make the conservative assumption that the only
source of non-adiabatic transitions are of the Landau-Zener (LZ)
type~\cite{Zener1932}, i.e., localized transitions between adjacent
levels at avoided crossings where the energy levels locally assume
the geometry of hyperbolae, then we can obtain a lower bound on the
transition probability using RMT. Additional failure modes can only
make the failure rate of the algorithm worse.

\begin{figure}[t]
\includegraphics[width=9cm]{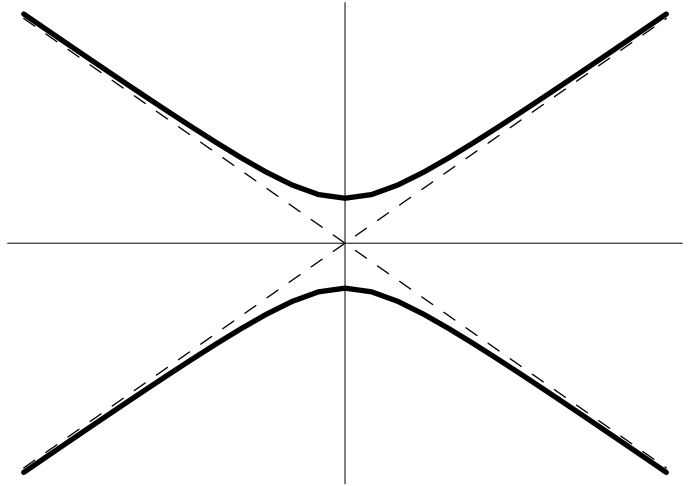}\\
\bigskip
\caption{Avoided level crossings can occur when adjacent levels
assume the geometry of converging hyperbolae. Both the slope
difference $\Delta m$ and minimum gap $\Delta E$ affect the
probability of a transition occurring.\label{fig1}}
\end{figure}

With these assumptions and the use of RMT, we can then determine
the transition rate from the ground state averaged over an
ensemble of problem instances having the same ratio of clauses to
variables.  There are two model-dependent quantities in this
result: the average ground state level spacing and the typical
size of LZ asymptotic slopes.  Ultimately, these quantities are
related to the parameters characterizing problem instances of the
type being solved, i.e., the ratio of clauses to variables. Hence
the transition rate, at a given point in the interpolation, is
related to the difficulty of the problem instances at that point.
The rest of the paper is organized as follows. Section II
describes the quantum adiabatic algorithm and the Landau-Zener
transition probability. Section III summarizes prior research on
the scaling properties of the algorithm, i.e., whether the
adiabatic criterion can be met if the interpolation is completed
in polynomial time. Section IV introduces the concepts of Random
Matrix Theory and Landau-Zener transitions needed for our
analysis. Section V reports on our numerical experiments and the
gap fluctuation phenomena they reveal. Section VI uses the
phenomena to justify a random matrix analysis of the adiabatic
algorithm and describes the implications on the scaling of the
quantum adiabatic algorithm. We discuss the distribution of gap
energies in an Appendix.

\section{The Adiabatic Algorithm}
The idea behind the quantum adiabatic algorithm is as follows. If
a quantum system is prepared in the ground state of a
time-independent Hamiltonian $H_0$, and if we then cause the
Hamiltonian to change from $H_0$ to a final form $H_1$ in $T$
steps, e.g., by driving it linearly
\begin{equation}
H(\frac{t}{T})=(1-\frac{t}{T})H_0 + \frac{t}{T}H_1\;,
\end{equation}
then the Adiabatic Theorem of Quantum Mechanics~\cite{Schiff1955}
guarantees that the system will remain in the ground state of the
instantaneous Hamiltonians $H(t)$, provided the change is made
sufficiently slowly, i.e., adiabatically. Thus, if the final
Hamiltonian can be made to encode a computational problem such
that the ground state of corresponds to the solution to this
problem, then the natural quantum mechanical evolution of the
system under the slowly changing Hamiltonian $H(t)$ would carry
our initial state into a final state corresponding to the
solution. A final state measurement would then reveal the
solution. The key question is how quickly can one drive the
interpolation between the initial and final Hamiltonians while
keeping the system in the ground state of the instantaneous
Hamiltonians passed through. If the shortest feasible
interpolation time scales polynomially with increasing problem
size, the quantum adiabatic algorithm would be deemed
``efficient", otherwise it would be deemed
``inefficient"~\cite{fn1}. An alternative way of stating this is
to ask under what conditions the passage from $H_0$ and $H_1$ can
be performed adiabatically~\cite{Schiff1955}. If the minimum
eigenvalue gap between the ground state $E_0$ and first excited
state $E_1$ of the instantaneous Hamiltonians is given by $g_{\rm
min}$, where
\begin{equation}
g_{\rm min} = \min_{0\leq t\leq T}\left[E_1(t)-E_0(t)\right]
\end{equation}
and the matrix element between the corresponding pair of
eigenstates is
\begin{equation}
\la\frac{dH}{dt}\ra_{1,0}=\la E_1;t|\frac{dH}{dt}|E_0;t\ra\;,
\ende
then the Adiabatic Theorem asserts that the final state will be
very close to the ground state of $H_1(T)$ , i.e.,
\begin{equation}
|\la E_0;T|\psi(T)\ra|^2\geq1-\epsilon^2
\end{equation}
provided that
\begin{equation}
\frac{|\la \frac{dH}{dt}\ra_{1,0}|}{g_{\rm min}^2}\leq \epsilon
\label{gap}
\end{equation}
where $\epsilon \ll 1$. If this criterion is met, we can be sure
the system will evolve into the desired state. But it is not
immediately clear how quickly we can interpolate between   and
while ensuring this adiabaticity criterion is not violated.

\section{Prior Analytic Results}
\label{sec3} Prior analytic studies of the adiabatic quantum
algorithm have yielded mixed results. Farhi et al. analyze several
models in which the gap behavior can be computed analytically, and
be shown to decrease polynomially in the problem
size~\cite{Farhietal2000}. However, they caution that the
particular problems they studied have a high degree of structure
that would also make them easy to solve classically. Nevertheless,
the results show that the adiabatic algorithm scales favorably at
least on easy problems. To show the gap is, at the very least,
non-vanishing, Ruskai~\cite{Ruskai2002} provides a clever proof
that the ground state of the instantaneous Hamiltonian must be
unique. However, as she points out, this tells us nothing about
the magnitude of the gap, and how it scales with problem size. A
less encouraging result was obtained by van Dam, Mosca, and
Vazirani~\cite{vandametal2001}, who were able to construct a
family of minimization problems for which they could prove an
exponential lower bound on the running time of the (original)
adiabatic algorithm on these problems. A subsequent paper by Farhi
et al.~\cite{Farhietal2002} challenged the inevitability of such
results by arguing that they might be circumvented by choosing a
different interpolation path between the initial Hamiltonian and
the one encoding the problem to be solved. To date, the most
sophisticated analysis of the running time of the adiabatic
algorithm on NP-complete problems was provided by Roland and
Cerf~\cite{RolandCerf2003}. They found that by nesting one quantum
adiabatic search algorithm within another, one can solve
NP-complete problems more efficiently than naive use of an
adiabatic version of Grover's algorithm. Nevertheless, the
run-time scaling is still exponential in problem size, albeit
better than what is possible classically.

\section{Applicability of Random Matrix Theory}

Random Matrix Theory is a statistical approach to Hamiltonian
systems that are otherwise analytically intractable. For example,
RMT focuses on universal model-independent properties of the
system under study, such as the distribution of the spectral {\it
fluctuations}. Many superficially different physical systems are
found to have distributions of spectral fluctuations that fall
into just a handful of categories. Such a characterization of
spectra originated in the context of nuclear
physics~\cite{Dyson1965} and was applied later to complex many
body systems and quantum systems having a chaotic classical
analog~\cite{Bohigas1991}. Once the distribution of spectral
fluctuations of a physical system has been identified (and deemed
to be irregular), RMT can be applied to make predictions about
properties of interest, such as transition rates between different
levels. For example, the problem of estimating transition rates
has been examined in the context of nuclear dissipation, and the
use of the LZ transition as a mechanism for nuclear dissipation
was suggested originally by Hill and
Wheeler~\cite{HillWheeler1952}. The combination of the LZ
transition probability with the RMT statistical approach was
examined by Wilkinson~\cite{Wilkinson1988,Wilkinson1990}, whose
results we apply to the current problem. Although the LZ
assumption is reasonable for adiabatic systems \cite{Sanchez1996},
the mechanism for dissipation in complex spectra continues to be
investigated in the context of RMT, with more recent approaches
using a non-LZ, propagator approach~\cite{Bulgacetal1996}.

To determine the applicability of RMT, the regularity of the entire
spectrum at each adiabatic interpolation point is measured by the
Brody parameter (defined below).  The nearest-neighbor spacing
distribution is the most common measure of spectral regularity in
quantum systems. This measure is quantified by the Brody parameter
$q$ that interpolates between a regular Poisson spectrum $(q=0)$ and
an irregular (quantum chaotic) Wigner distribution
$(q=1)$~\cite{Brody1973}. A renormalized spectrum with a Brody
parameter $q$  is characterized by the following NNS probability
distribution for spacing level spacing $\delta$:
\begin{equation}
p_q(\delta)=(1+q)\beta\delta^q\exp({-\beta\delta^{1+q})}\;,\;
\beta=\left[\Gamma\left(\frac{2+q}{1+q}\right)\right]^{1+q}\;.\label{fit}
\end{equation}
The form of this distribution, for different values of the Brody
parameter, is shown in Fig.~\ref{Fig2}.

\begin{figure}
\includegraphics[width=8cm]{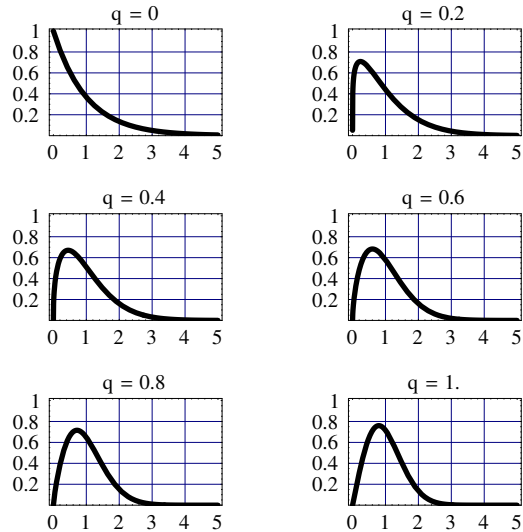}\\
  \caption{Nearest neighbor probability distribution $p_q(\delta)$ for eigenvalue spacings $\delta$,
  in unfolded spectra for values of the Brody parameter $q\in [0,1]$. When $q=0$, the distribution
  resembles an exponential distribution, but changes to unimodal as $q$ is
  increased.
  }\label{Fig2}
\end{figure}
Irregular RMT spectra are characterized by an abundance of avoided
level crossings, and a
lack of level degeneracies. We now proceed to determine whether or
not the distribution of spectral fluctuations is anywhere
irregular during the interpolation process.

\section{Spectral Fluctuation Experiments}
Our first task is to determine an appropriate ensemble of random
soluble 3-SAT problem instances to use. We would like to use
computationally hard problem instances, because we are most
interested in assessing the scaling of the failure rate of the
adiabatic algorithm on hard problems. Hard, in this sense, is a
relative term. When solving random instances of soluble 3-SAT
problems having $n$ variables and $m$ clauses, typically the
hardest instances are encountered at a critical value of the
clause to variable ratio $m/n$. For the 3-SAT problem as
$n\to\infty$, the hardest instances are clustered around the ratio
$m/n\approx4.2$. However, small problem instances (having, say, $n
< 40$), typically have a somewhat displaced transition point.
Fig.~\ref{Fig3}. shows the mean computational cost of solving
3-SAT problems containing from $n = 8$ to $n = 50$ variables using
either the Davis-Putnam (DP) algorithm~\cite{Davisetal1962} or the
GSAT algorithm~\cite{Selmanetal1992}. Regardless of the algorithm
used, an easy-hard-easy pattern is apparent when the number of
clauses is increased at fixed number of variables. In the limit of
infinite problem size, the easy and hard instances are separated
by a phase transition (see, e.g.,~\cite{KirkpatrickSelman1994}).
But, as can be seen from Fig.~\ref{Fig3}, the location of the
phase transition point is extremely ill-defined for problem
instances having $n < 20$. Thus inferring any reliable
cost-scaling by extrapolating costs from such small instances
would be exceedingly unreliable. Yet this is exactly what has been
done in assessing the scaling of the quantum adiabatic algorithm
from numerical experiments~\cite{Farhietal2001}. As the simulation
of the adiabatic algorithm solving an $n$ variable 3-SAT problem
involves $2^n\times2^n$-dimensional matrices, we cannot simulate
very large cases. Hence, we begin by first determining the {\em
actual} location of the hardest problems for 3-SAT problems
involving a more tractable $n=8$ variables, rather than relying on
the asymptotically known result whose applicability is suspect at
small $n$. Specifically, we generated 72,000 random 3-SAT problem
instances all having $n=8$ variables, but with the number of
clauses ranging from $m=8$ to 80, corresponding to $1\le m/n \le
10$, and solved them using the GSAT algorithm. Each data point was
computed from an average of 1000 problem instances. The results
are shown in the lower portion of Fig.~\ref{Fig3}. Although the
difference between easy cases and hard cases is not as pronounced
as it is for much larger $n$, nevertheless, the data suggests that
problem instances centered around $m/n=6$  will yield relatively
hard cases~\cite{fn3}. We will use such problem instances to
create the ensemble we need in our numerical studies of the
distribution of spectral fluctuations of the instantaneous
Hamiltonians encountered during the interpolation phase of the
adiabatic algorithm.

\begin{figure}
%
\includegraphics[width=6cm]{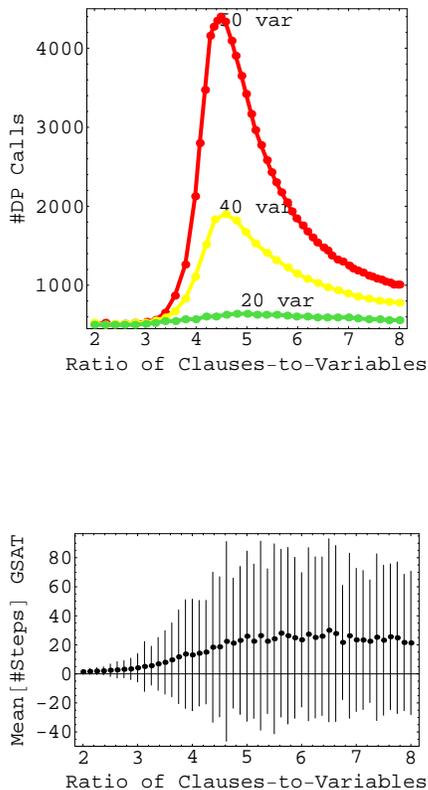}\\
\caption{Upper plot shows the mean computational cost of solving
$n = 20, 40,$ and 50 variable 3-SAT problems using the
Davis-Putnam algorithm (the data on the scaling of this algorithm
is due to Bart Selman, see~\cite{KirkpatrickSelman1994}). The data
show that for small problems (i.e., $n<50$)  the phase transition
region is smeared out. Lower plot shows the mean computational
cost of solving 1000 random instances of guaranteed soluble 3-SAT
problems having $n$ = 8 variables, and $m$ = 8 to 80 clauses,
using the GSAT algorithm. Again, for such small-sized problems the
region of hard problem instances (relative to other instances) is
quite spread out, but roughly centered on $m/n = 6$, rather than
the asymptotic value of 4.2.\label{Fig3}}
\end{figure}

%
Next we turn our attention to the global spectral properties of
the instantaneous Hamiltonians encountered in the quantum
adiabatic algorithm for easy and hard problem instances.
Specifically, we obtain the NNS distribution of the adiabatic
Hamiltonian $H(s)$ after renormalization of the spectrum to unit
average local level density~\cite{Mehta1991,Brodyetal1981} for an
ensemble of easy cases, and for an ensemble of hard cases.  The
regularity of the spectrum is determined at each point in the
adiabatic evolution by fitting it to (\ref{fit}), and obtaining
the Brody parameter $q(s)$ of the NNS distribution at that point.
We begin by determining the spectral distribution of an ensemble
of easy problems as the interpolation parameter in the quantum
adiabatic algorithm ranges from $s$ = 0 to $s$ = 1, for instances
of soluble 3-SAT having $n=8$ variables and $m=4$ clauses, i.e.,
problems for which $m/n = 0.5$. Each histogram in Fig.~\ref{Fig4}
is based on the spectral behavior of an ensemble of 20 problem
instances having fixed values of $n$ and $m$. For the easy
problems, the spectrum of each instantaneous Hamiltonian conforms
to a Poisson (regular) spectral fluctuation distribution, and
small Brody parameter, $q$ = 0. Hence, we conclude that RMT is not
applicable in the easy region, i.e., for $m/n<5$ ($n$=8).

\begin{figure}
\includegraphics[width=9cm]{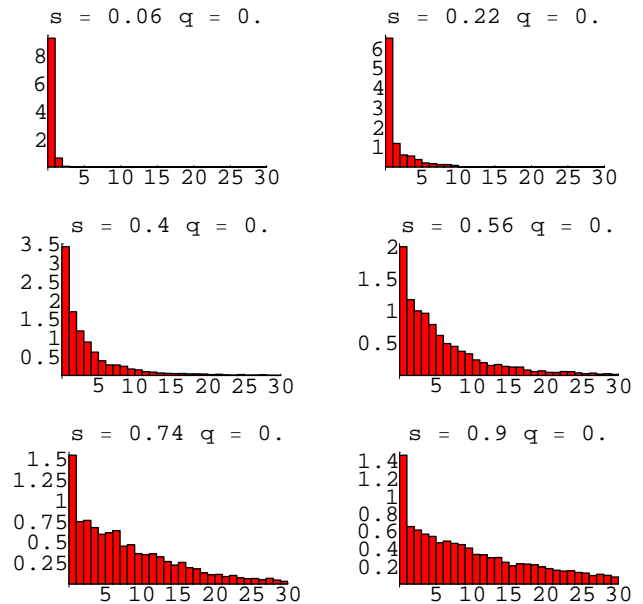}\\
\caption{Distribution of eigenvalue gaps in the unfolded spectra
of the instantaneous Hamiltonians during the course of the
adiabatic algorithm solving easy instances of 3-SAT having $n = 8$
variables, and $m = 4$ clauses, as the interpolation parameter
varies from 0 to 1. As $s$ increases, the Brody parameter remains
zero, implying that RMT is not applicable to easy problem
instances.  The horizontal axes are nearest neighbor eigenvalue
spacings (NNS), expressed in tenths of the mean local
spacing.}\label{Fig4}
\end{figure}

In contrast, Fig.~\ref{Fig5} shows the eigenvalue gap fluctuations
of instantaneous Hamiltonians induced from ``hard" instances of
soluble 3-SAT with $n=8$ variables and $m=48$ clauses. Again, our
ensemble averages over 20 instances of a fixed $n$ and $m$. For
$m/n=6$ (hard problems) we observe Poisson behavior for $s<0.5$,
but for $s> 0.5$ the spectra become increasingly irregular and the
Brody parameter becomes significant.
In other words, the instantaneous Hamiltonians induced by random,
hard 3-SAT instances, appear to have a qualitatively different
spectrum from those of easy problems of the same size. In
particular, at a certain point in the interpolation process
between the initial and final Hamiltonians, the spectrum becomes
irregular, and the NNS distribution resembles a Wigner
distribution with a relatively large Brody parameter. Here, RMT
can be applied to estimate transition rates between levels.

Finally, one can repeat these experiments for problems having $n$
= 8 variables at a clause to variable ratio of $m/n=9$ (easy
region again - data not shown). Here, we observe results similar
to $m/n=6$, but with a slightly smaller maximum Brody parameter.
We attribute this to the fact that $m/n=9$ is easier to solve than
$m/n=6$, although for $n$ = 8 variables, this difference is
slight.
\begin{figure}[t]
\includegraphics[width=9cm]{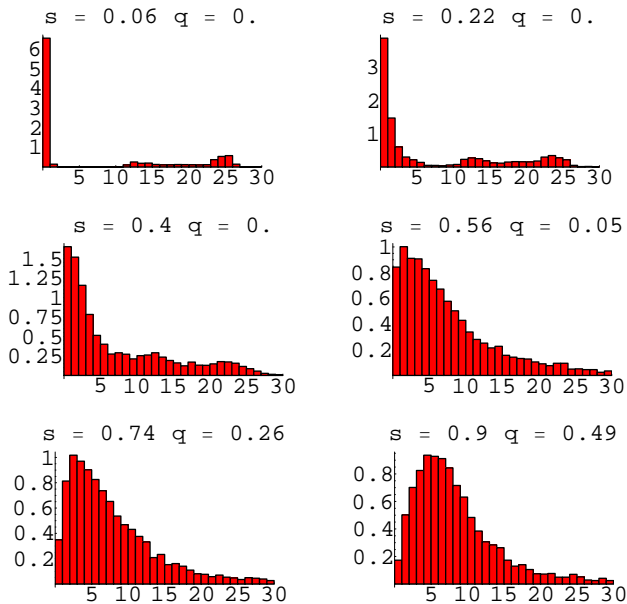}\\
\caption{Distribution of eigenvalue gaps in the unfolded spectra
of the instantaneous Hamiltonians during the course of the
adiabatic algorithm solving hard instances of 3-SAT having $n$ = 8
variables, and $m$ = 48 clauses, as the interpolation parameter
varies from 0 to 1. As $s$ increases, the Brody parameter varies
from 0 to 0.49, and back to 0.  The horizontal axes are nearest
neighbor eigenvalue spacings (NNS), expressed in tenths of the
mean local spacing. }\label{Fig5}
\end{figure}

\begin{figure}
\includegraphics[width=9cm]{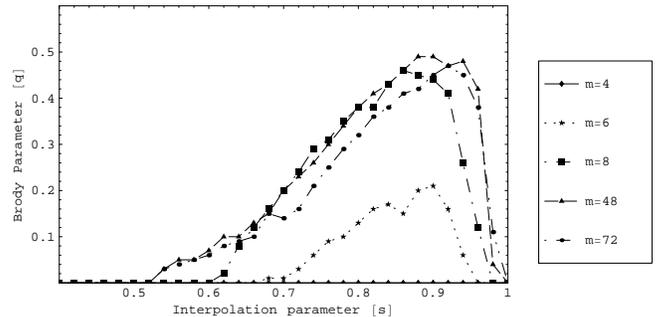}\\
\caption{Brody parameter as a function of the interpolation
parameter s,for easy and hard instances of 3-SAT with $n$ = 8
variables  from $m$ = 4 to 72 clauses. Note that the largest
values of the Brody parameter coincides with the hardest problems.
For soluble 3-SAT problems of size $n$ = 8, these are found around
$m/n$ = 6 rather than at $m/n = 4.2$, which is the asymptotically
valid transition point for 3-SAT as $n\to\infty$.}\label{Fig6}
\end{figure}

Fig.~\ref{Fig6} summarizes the Brody parameter as a function of
the interpolation parameter $s$ for easy and hard problems with
$n=8$ variables, and from 4 to 72 clauses. The critical question
in deciding if the quantum adiabatic algorithm can be completed in
polynomial time is whether such a (fast) interpolation would
induce level transitions. If they do occur, the system will not
reside in the ground state of $H_1$ upon completion of the
adiabatic path, and the algorithm will  have failed to find the
solution. In the next section, we compute the probability that
level transitions occur during the course of the interpolation
from  $H_0$ to $H_1$ using RMT, for problems of a given degree of
difficulty.

\section{Random Matrix Analysis of Quantum Adiabatic Algorithm}

Previous analyses of the adiabatic algorithm (summarized in
Section \ref{sec3}) placed the greatest significance on the
scaling of the $E_1-E_0$ gap with increasing problem size.
However, as Equation (\ref{gap}) shows, it is the ratio of the
matrix element $\la dH/dt\ra_{1,0}$ to the square of the minimum
gap that determines whether the adiabatic theorem applies. Here,
we calculate instead the {\em probability of a transition} from
the ground state, which is a reliable proxy for the failure rate
of the adiabatic algorithm. In regions where the Brody parameter
is significant, we assume that any non-adiabatic transitions are
of Landau-Zener type, i.e., confined to adjacent levels at avoided
crossings where the energy levels assume the geometry of
convergent hyperbolae~\cite{Zener1932} (see Fig.~\ref{fig1}). When
approached in this manner, the probability for a single
transition, anywhere in the spectrum, can be parameterized by the
minimum gap size $\Delta E$, the difference in the asymptotic
slopes $\Delta m$, and the rate of change of the adiabatic
evolution parameter $ds/dt=\dot s$. Specifically, the transition
probability is~\cite{Zener1932}
\begin{equation}
P=e^{-2\pi\gamma}\;,\;\; \gamma=\frac{1}{4\hbar}\frac{\Delta
E}{|\Delta m|\dot s}\;.
\end{equation}
Typical values for the parameters $\Delta E$  and $\Delta m$ will
vary with the difficulty of the problem instance being solved
(reflected by the clause to variable ratio $m/n$) as well as with
the interpolation parameter.

In order to exploit this transition probability to predict the rate
of transition from the ground state, we need to verify that the
$E_0-E_1$ gap fluctuations follow the same distribution as those in
the body of the spectrum. In Appendix A, we show that the
distribution of gap fluctuations is characterized by a Brody
parameter comparable to the typical fluctuations in the body of the
spectrum, establishing this point. To incorporate the model
dependence (non-universality) of the problem, the average local
level density of the spectral region of interest (here the levels)
is included.  Thus, the RMT/LZ calculation results in an
ensemble-averaged transition rate with the LZ process as the
principle transition mechanism.

The transition rate from the $i$th eigenstate of the instantaneous
Hamiltonian with energy $E_i$, $dP_i/dt$, can be written in terms of
the second moment of the occupation probability distribution
$P_i(t)$~\cite{Wilkinson1988} because it essentially is the rate of
{\em diffusion} of the occupation probability.
As discussed in the introduction, the appropriate RMT ensemble to
use for physical systems with exponential densities of states is the
embedded GOE, consisting of real, orthogonal matrices having
Gaussian-distributed random matrix elements with at most two-body
interactions. Since in the calculation of transition rates in RMT
only the fluctuation properties of the spectrum (rather than the
density of states proper) enter~\cite{Wilkinson1988}, we can safely
substitute a GOE to obtain the probability to transition from the
ground state
\begin{equation}
\frac{dP_0}{dt}\propto \sigma^{3/2}\rho^2|\dot s|^{3/2}\;,
\label{rate}
\end{equation}
where $\rho$ is the level density of $E_0-E_1$ levels averaged over
an ensemble of problem instances having the same clause to variable
ratio, and $\sigma$ is the typical size of the asymptotic slopes of
the LZ avoided level crossings in the region of interest. Indeed, as
shown in Ref.~\cite{BenetWeidenmueller2003}, embedded GOEs give rise
to NNS distributions very similar to those arising in a GOE, except
that the maximal Brody parameter is limited to $q\approx 0.8$. Note
that this is also implied by a {\em numerical}
calculation~\cite{Sanchez1996} of the dissipation rate of
probability distributions of physical instantaneous Hamiltonians
(i.e., those drawn from and embedded GOE), which agree with the
scaling in Eq.~(\ref{rate}).

We emphasize that Eq.~\ref{rate} expresses the transition rate from
the ground state in terms of the average $E_0-E_1$ level density,
rather than the minimum gap (i.e., the maximum level density). The
average level density is the more relevant parameter for assessing
the typical behavior of the adiabatic algorithm. As the adiabatic
algorithm fails if the system transitions from the ground state,
Eq.~\ref{rate} can be interpreted as the average failure rate of the
adiabatic algorithm when driven with uniform interpolation velocity
$ds/dt$.

Finally, we need to determine the scaling of the transition rate
from the ground state (i.e., the failure rate) with increasing
problem size. We do this in two stages: first we show how the
transition rate must scale with $\rho$ in order to keep the
transition rate bounded, then we show how $\rho$ scales with
problem size. Although the values of $\rho$ and $\sigma$ are
model-dependent quantities, we note that $\sigma$ is the
characteristic slope in the energy-parameter space (i.e.,
$\sigma\approx dE/ds$). To make explicit the dependence on level
density, we work with unfolded energies
($E_i=\Delta\times\epsilon_i$) where $\Delta$ is the mean level
spacing ($\Delta=1/\rho$) and $\epsilon_i$ are the unfolded energy
levels having mean level spacing one. Under this transformation,
$\sigma^{3/2}\rightarrow(\Delta\times\tilde\sigma)^{3/2}$, and we
therefore write
\begin{equation}
\frac{dP}{dt}=\propto\tilde \sigma^{3/2}\rho^{1/2}|\dot s|^{3/2}\;.
\label{unfold}
\end{equation}

In general, while the unfolded level slope $\tilde\sigma$  and the
average level density $\rho$ are model-dependent
quantities~\cite{Mitchelletal1996}, Eq.~(\ref{unfold}) nevertheless
exhibits the explicit dependence of the transition rate on $\rho$.
We can establish a lower bound on the interpolation time required to
evolve the system through the irregular region by noting first that
\begin{equation}
\frac{dP}{dt}(\dot s,\rho)\geq \frac{dP}{dt}(\dot s, \rho_{\rm
min})
\end{equation}
where we defined $\rho_{\rm min}=\min_s \rho(s)$ and where the
minimization is carried out over only those values of $s$ that
give rise to an irregular spectrum. It is important to note that
$\rho_{\rm min}$ is not the minimum of a particular problem
instance; it is an average level density at a particular value of
$s$, whose value is constant for a given problem parameter set.
The failure probability during an evolution path of length $T$ is
then bounded by
\begin{equation}
P\geq {\rm const} \tilde \sigma^{3/2}\Delta
s^{3/2}\left[\frac{\rho_{\rm min}}{T}\right]^{1/2}\;. \label{prob}
\end{equation}

To ensure a given transition probability over a given range
$\Delta s$ in an irregular region, the interpolation time $T$ must
scale as $\rho_{\rm min}$. If we can now estimate how $\rho_{\rm
min}$ (an average quantity) scales with increasing problem size,
we can estimate how the time needed to complete a particular part
of the adiabatic algorithm must scale in order to keep the
transition rate from the ground state small. This part is
precisely the region where one would need to go most slowly to
avoid an unwanted transition from the ground state. What can we
say about the scaling of $\rho_{\rm min}$ with problem size? As
pointed out by Ruskai~\cite{Ruskai2002}, in regions characterized
by a lack of level degeneracies (such as irregular spectral
regions) the interpolating Hamiltonian $H(s)$  must fit $2^n$
eigenvalues into a range that is polynomial in $n$, and
consequently the level density must scale {\em exponentially} with
problem size. Our numerical simulations provide strong evidence
for the existence of such irregular spectral regions by the fact
that we obtain significant values of the Brody parameter for
distributions averaged over instances that have a sizable degree
of difficulty (as measured by their clause to variable ratio. In
other words, difficult problem instances display spectra
characterized by a large Brody parameter and few level
degeneracies, and therefore show a lack of symmetry. Thus, with
the final assumption that non-degenerate spectra (irregular
spectra having a large Brody parameter) have a minimum average
level density that scales exponentially with problem size, we find
the transition rate (\ref{prob}) out of the ground state (the
algorithmic failure rate), and thus that the adiabatic evolution
time cannot scale polynomially with problem size.

The information regarding the parameter regions having significant
transition probability may be used in a variable-speed approach to
the adiabatic quantum algorithm in a manner similar to
\cite{RolandCerf2003}.  In this way, a speedup over classical
algorithms for solving NP-complete problems still appears
possible. We wish to emphasize that Eq.~\ref{prob}, and the entire
RMT analysis, holds only in those regions of the interpolation
process where we find irregular spectra, i.e., non-trivial Brody
parameters. In particular, the formula does not apply to regions
of the interpolation having a small Brody parameter, arising
typically in the very early and very late stages of a hard
problem, or at all stages of an easy problem. Therefore, our
formula is not inconsistent with previous results (e.g., in
Section~\ref{sec3} showing that the adiabatic algorithm scales
favorably on easy problems), because in such cases
Eq.~(\ref{prob}) is invalid.

\section{Conclusion}

Our approach to the analysis of Adiabatic Quantum Computation is
statistical in nature, and relies on the applicability of Random
Matrix Theory to large Hamiltonians.  With such an approach, we
examine the behavior of an averaged failure rate and interpolation
time when averaged over an ensemble of problem instances of a
given difficulty.  We find a spectral transition from orderly to
disorderly as problem difficulty is increased.  Our analysis of
these results suggests using the Landau-Zener transition
probability in a statistical RMT approach.  For those regions
where RMT applies, degeneracies are lacking and we show that
average failure rate, and average interpolation time do not scale
polynomially with problem size. Nevertheless, the quantum
adiabatic algorithm appears to be more efficient than any known
classical algorithm for solving NP-complete problems, with a
speed-up commensurate with Grover's search algorithm.  Moreover,
the insights we gain from the spectral fluctuation analysis may be
used to design a {\it variable} interpolation rate, which has the
potential to improve the adiabatic algorithm over its performance
using a uniform interpolation rate.

\begin{figure}[t]
\includegraphics[width=9cm]{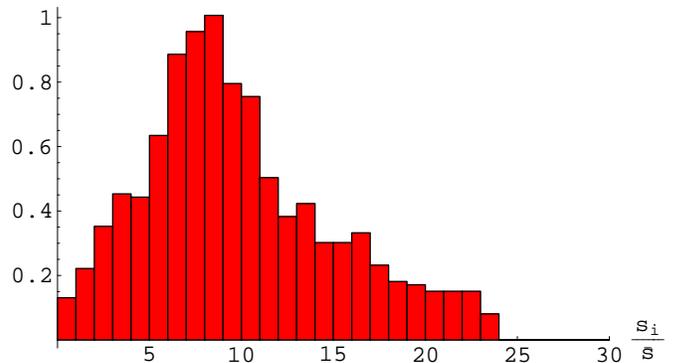}\\
\caption{NNS distribution of $E_0-E_1$ gaps for 1000 random
problem instances in the hard region for problems of size $n$ = 8.
The horizontal axis is the eigenvalue spacing in tenths of the
mean level spacing.}\label{Fig7}
\end{figure}

Finally, we believe the connection shown here between difficult
computational problems and physical spectral irregularity is a
powerful path towards understanding and classifying physical
approaches to computation.


\appendix

\section*{Appendix}

Figure 7 shows the ground state level distribution of $E_0-E_1$ gaps
for 1000 random problem instances in the hard region for problems of
size $n$ = 8. The horizontal axis is the eigenvalue spacing in
tenths of the mean level spacing.  The distribution shows that
spectral irregularity definitely extends to the lowest lying levels
for hard instances of 3-SAT problems, implying that RMT is
applicable to hard instances of 3-SAT across the entire spectrum of
levels, including the $E_0-E_1$ gap, which is the relevant gap for
assessing the scaling of the quantum adiabatic algorithm. Therefore,
our use of RMT to characterize the transition rate out of the ground
state, and hence to estimate the cost scaling of the quantum
adiabatic algorithm, is valid.

\vskip 0.5cm {\bf Acknowledgement} The research described in this
paper was performed at the Jet Propulsion Laboratory (JPL),
California Institute of Technology, under contract with the
National Aeronautics and Space Administration (NASA).  We thank
the JPL Supercomputing Project for the use of the Cray
supercomputer used in computations. DM received fellowship support
through the NASA Faculty Fellowship Program (NFFP). CPW thanks the
Advanced Research and Development Activity and the National
Security Agency for support. CA is supported by the Army Research
Office.


\begin{thebibliography}{9}

\bibitem{Bennettetal1997}
C.H. Bennett, E. Bernstein, G. Brassard, and U. Vazirani,
Strengths and weaknesses of quantum computing. SIAM J. Computing
(spec. issue), 1997.

\bibitem{Farhietal2001}
E. Farhi, J. Goldstone, S. Gutmann, J. Lapan, A. Lundgren, and D.
Preda. A quantum adiabatic evolution algorithm applied to random
instances of an NP-complete problem, Science {\bf 292}, 472-475
(2002).

\bibitem{GareyJohnson1979}
M. R. Garey and D. S. Johnson, {\it Computers and Intractability:
A Guide to the Theory of NP-Completeness}. \newblock (W.\ H.\
Freeman \& Company, 1979).

\bibitem{Selmanetal1992}
B. Selman, H. J. Levesque, and D. G. Mitchell, A new method for
solving hard satisfiability problems, in {\it Proceedings of the
Tenth National Conference on Artificial Intelligence (AAAI-92)},
440-446. (AAAI Press/MIT Press, 1992).

\bibitem{LynceMarquesSilva2002}
I. Lynce and J. Marques-Silva, Building state-of-the-art SAT
solvers, In {\it Proc. of the European Conference on Artificial
Intelligence} (IOS Press, Amsterdam, 2002).

\bibitem{Mehta1991}
M.L. Mehta, {\it Random Matrices}, (Academic Press, San Diego,
1991).

\bibitem{Wigner1967}
E.P. Wigner, Random matrices in physics, SIAM Review {\bf 9}, 1-23
(1967).

\bibitem{Brodyetal1981}
T. A. Brody, J. Flores, J. B. French, P. A. Mello, A. Pandey, and
S. S. M. Wong, Random matrix physics: Spectrum and strength
fluctuations, Reviews of Modern Physics {\bf 53}, 385 (1981).

\bibitem{MonFrench1975}
K. K. Mon and J.B. French, Statistical properties of many-body
spectra, Ann. Phys. (N.Y.) {\bf 95}, 90-111 (1975).

\bibitem{Brody1973}
T. A. Brody, Statistical measure for repulsion of energy levels,
Lett. Nuovo Cimento {\bf 7}, 482 (1973).

\bibitem{Zener1932}
G. Zener, Non-adiabatic crossing of energy levels, Proc. Roy. Soc.
{\bf A 137}, 696 (1932).

\bibitem{Schiff1955}
L. Schiff, {\it Quantum Mechanics}, (McGraw-Hill, New York, 1955).

\bibitem{fn1}Note that, even if the quantum adiabatic algorithm proves to be
inefficient by this measure it may, nevertheless, be superior to
any  known classical algorithm.

\bibitem{Farhietal2000}
E. Farhi, J. Goldstone, S. Gutmann, and M. Sipser, Quantum
computation by adiabatic evolution, quant-ph/0001106 (2000).

\bibitem{Ruskai2002}
M. B. Ruskai, Comments on adiabatic quantum algorithms,
Contemporary Mathematics {\bf 307}, 265-274 (2002).

\bibitem{vandametal2001}
W. van Dam, M. Mosca, and U. Vazirani, How powerful is adiabatic
quantum computation?, in {\it Proceedings of the 42nd Annual
Symposium on Foundations of Computer Science}, pp. 279-287 (2001).

\bibitem{Farhietal2002}
E. Farhi, J. Goldstone, and S. Gutmann, Quantum adiabatic
evolution algorithms with different paths, quant-ph/0208135
(2002).

\bibitem{RolandCerf2003}
J. Roland and N. J. Cerf, Adiabatic quantum search algorithm for
structured problems, Phys. Rev. {\bf A 68}, 062312 (2003).

\bibitem{Dyson1965}
F. J. Dyson, in {\it Statistical Theories of Spectra:
Fluctuations}, C. E Porter, ed., (Academic Press, New York, 1965).

\bibitem{Bohigas1991}
O. Bohigas, Random matrices and chaotic dynamics, in {\it Chaos
and Quantum Physics}, M. Giannoni, A. Voros, and J. Zinn-Justin,
eds. (North-Holland, New York, 1991), pp. 87-199.

\bibitem{HillWheeler1952}
D.L. Hill and J.A. Wheeler, Phys. Rev. {\bf 89}, 1102 (1952).

\bibitem{Wilkinson1988}
M. Wilkinson, Statistical aspects of dissipation by Landau-Zener
transitions, J. Phys. {\bf A 21}, 4021 (1988).

\bibitem{Wilkinson1990}
M. Wilkinson, Diffusion and dissipation in complex quantum
systems, Phys. Rev. {\bf A 41}, 4645 (1990).

\bibitem{BenetWeidenmueller2003}
L. Benet and H. A. Weidenmuller, Review of the k-body embedded ensembles of Gaussian random matrices, J. Phys. {\bf A 36}, 3569-3593 (2003).

\bibitem{Sanchez1996}
M.J. Sanchez, E. Vergini, and D. A. Wisniacki, Characterization of
Landau-Zener transitions in systems with complex spectra, Phys.
Rev. {\bf E 54}, 4812 (1996).

\bibitem{Bulgacetal1996}
A. Bulgac, G. DoDang, and D. Kusnezov, Random matrix approach to
quantum dissipation, Phys. Rev. {\bf E 54}, 3468 (1996).

\bibitem{Mitchelletal1996}
D. Mitchell, Y. Alhassid, and D. Kusnezov, Gaussian process and
universal parametric decorrelations of wavefunctions, Phys. Lett.
{\bf A 215}, 21 (1996).

\bibitem{Davisetal1962}
M. Davis, G. Logemann, and D. Loveland, A machine program for
theorem proving, Comm. ACM {\bf 5}, 394 (1962).

\bibitem{KirkpatrickSelman1994}
S. Kirkpatrick and B. Selman, Critical behavior in the
satisfiability of random Boolean expresssions, Science {\bf 264},
1297-1301 (1994).

\bibitem{fn3}It is not our intention to use simulation results to infer the
cost scaling of the adiabatic algorithm. Rather, we use
simulations merely to infer the applicability of a particular
mathematical technique---random matrix theory. Hence, while using
problem instances with    variables is not ideal,  it is unlikely
that we gain much additional insight by using (say) instances
having n = 20 rather than n = 8, because (as Fig.~3 shows) even at
n = 20 we would still be grappling with an anomalously broad phase
transition phenomenon.



\end{thebibliography}
\end{document}